\documentstyle[12pt]{article}

\newcommand{\newsection}{    
\setcounter{equation}{0}
\section}
\newcommand{\tr}[1]{\,{\rm tr}\,#1}
\newcommand{\ntr}[1]{\,\frac {\rm tr}{N}\,#1}
\def\e{{\,\rm e}\,}

\def\eop{\vspace*{\fill}\pagebreak}
\def\be{\begin{equation}}
\def\ee{\end{equation}}
\def\bea{\begin{eqnarray}}
\def\eea{\end{eqnarray}}
\def\LA{\left\langle}
\def\RA{\right\rangle}
\newcommand{\Di}{\,\hbox{Disc}_\nu\,}
\newcommand{\Co}{\,\hbox{Cont}_\nu\,}
\newcommand{\rf}[1]{(\ref{#1})}
\newcommand{\eq}[1]{Eq.~(\ref{#1})}

\def\a{\alpha}

\def\L{\Lambda}
\def\l{\lambda}
\def\om{\omega}
\newcommand{\ie}{{\it i.e.}\ }
\newcommand{\half}{{\textstyle{1\over 2}}}
\newcommand{\p}{{\prime}}
\newcommand{\ra}{\rightarrow}
\newcommand{\ci}{\int_{C_1}\frac{d\omega}{2\pi i}}
\newcommand{\cii}{\int_{C_2}\frac{d\omega}{2\pi i}}

\newcommand{\eps}{\varepsilon}

\newcommand{\non}{\nonumber \\*}
\let\la=\lambda

\newcommand{\VVp}{{\cal V}^\prime}

\newcommand{\real}{\,\hbox{Re}\,}
\newcommand{\im}{\,\hbox{Im}\,}
\def\laa{\mathrel{\mathpalette\fun <}}

\def\fun#1#2{\lower3.6pt\vbox{\baselineskip0pt\lineskip.9pt
\ialign{$\mathsurround=0pt#1\hfil##\hfil$\crcr#2\crcr\sim\crcr}}}

\begin{document}

\begin{titlepage}
\begin{flushright}
NBI-HE-93-28 \\ May, 1993
\end{flushright}
\vspace{1cm}

\begin{center}
{\LARGE Some Remarks about the Two-Matrix Penner}  \\ \vspace{0.6cm}
{\LARGE  Model and the Kazakov--Migdal Model} \end{center} \vspace{.5cm}
\begin{center}
{\large Yu.\ Makeenko}\footnote{E--mail: \ makeenko@nbivax.nbi.dk \ / \
makeenko@desyvax.bitnet \ / \ makeenko@vxitep.itep.msk.su \ }
\\ \mbox{} \\
{\it The Niels Bohr Institute, 2100 Copenhagen, DK} \\ \vskip .2 cm
and  \\  \vskip .2 cm
{\it Institute of Theoretical and Experimental Physics, 
117259 Moscow, RF}
\end{center}

\vskip 1.5 cm
\begin{abstract}
I consider the Hermitean two-matrix model with a logarithmic potential
which is associated in the one-matrix case with the Penner model.
Using loop equations I find an explicit solution of the model
at large $N$ (or in the spherical approximation) and demonstrate  that
it solves the corresponding Riemann--Hilbert problem.
I construct the potential of the Kazakov--Migdal model on a $D$-dimensional
lattice, which turns out to be a sum of two logarithms as well, whose large-$N$
solution is given by the same formulas.
In the ``naive'' continuum limit this potential recovers in \mbox{$D<4$}
dimensions the standard scalar theory with quartic self-interaction.
I~exploit the solution to calculate explicitly the pair correlator of gauge
fields in the Kazakov--Migdal model with the logarithmic potential.

\end{abstract}

\vspace{2cm}
\noindent
Submitted to {\sl  Physics Letters B}

\eop
\end{titlepage}
\setcounter{page}{2}

\section{Introduction}

The Penner model~\cite{Pen86} is the Hermitean {\it one-matrix}\/ model
defined by the partition function
\be
Z_1=\int d\phi \e^{-N \tr{V(\phi)}}
\label{1mamo}
\ee
with the logarithmic potential
\be
V(\phi)=-\a \Big(\log{(1+\phi)} -\phi\Big) \,.
\label{penpot}
\ee
This model has been extensively studied in the modern context of matrix
models after the observation by Distler and Vafa~\cite{DV91} that the
partition function~\rf{1mamo} with the potential~\rf{penpot} coincides
with that of the compactified one-dimensional string at a self-dual
radius. Some generalizations of the Penner model~\cite{Tan91}
were considered and solved in the large-$N$ limit. Of special interest
is the fact~\cite{CM92} that an external field problem for the Penner
model is equivalent to the Hermitean one-matrix model~\rf{1mamo} with a
general potential $V(\phi)$ which is analytic at $\phi=0$.

In the present paper I consider the Hermitean {\it two-matrix}\/ model
\be
Z_2=\int d\phi_1 d\phi_2 \e^{N\tr{}\Big(\phi_1\phi_2-{\cal V}(\phi_1)
-{\cal V}(\phi_2) \Big)}
\label{2mamo}
\ee
with the Penner potential~\rf{penpot}. One of the motivations is the
recent paper~\cite{DKK93} where the two-dimensional minimal conformal
field theories have been constructed from the two-matrix
model~\rf{2mamo} with  polynomial potentials as was first conjectured by
Douglas~\cite{Dou90}.
{}From this point of view it
is interesting to find out what happens for the {\it non-polynomial}\/
potential~\rf{penpot}.

Another motivation is that the two-matrix model~\rf{2mamo} describes the
solution of the Kazakov--Migdal~\cite{KM92} model
which is defined by the partition function
\be
Z_{KM}=\int \prod_{x,\mu} dU_{\mu}(x) \prod_x
 d\phi_x \e^{ \sum_x N \tr{}\left(-V(\phi_x)+
\sum_{\mu=1}^D\phi_x U_\mu(x)\phi_{x+\mu}U_\mu^\dagger(x)\right)}\,,
\label{spartition}
\ee
where the integration over the gauge field $U_\mu(x)$ is over the Haar measure
on $SU(N)$ at each link of a $D$-dimensional lattice with $x$ labeling
its sites.
The Kazakov--Migdal model is an interesting extension of the matrix
models~\rf{1mamo} and \rf{2mamo} to the multi-dimensional case
and obviously recovers the two-matrix model if the lattice is just
two-points since the gauge field can be absorbed by a unitary
transformation, say of $\phi_1$.

While the fact that the models~\rf{2mamo} and \rf{spartition} are
equivalent at large $N$ is known~\cite{Kaz92},
an explicit relation between the potentials ${\cal V}$
in~\rf{2mamo} and $V$ in~\rf{spartition} has been studied
recently~\cite{DMS93}. The idea of the DMS approach is first to solve
the two-matrix model~\rf{2mamo}
in the large-$N$ limit for some choice of ${\cal
V}$ which is practically convenient and then, having the solution, to
calculate the associated potential $V$ of the Kazakov--Migdal model.
To realize this program, the loop equation for the one-link correlator
was derived whose solution completely determines all the correlators of
the model, in particular, the pair correlator of the gauge fields. Besides
the Gaussian case when both potentials are quadratic,
some explicit relations between ${\cal V}$ and $V$ have been studied.

In the present paper I show that for the Penner potential~\rf{penpot}
(in fact for a little bit more general one given by \eq{ttV} below) the
associated potential $V$ of the Kazakov--Migdal model is given by the sum of
two logarithms as well (\eq{V} below) so that the model~\rf{spartition}
with this potential is explicitly solved at large $N$ for
any $D$. The solution is
conveniently described via the one-matrix model whose potential (\eq{tV}
below) coincides with $V$ of the Kazakov--Migdal model for $D=0$.
I demonstrate that
this solution satisfies the master field equation derived by
Migdal~\cite{Mig92a} and recovers by turning the parameters of the potential
the solution for the quadratic potential found by Gross~\cite{Gro92}.
I study some properties of the Kazakov--Migdal model with this
logarithmic potential and show that in the ``naive'' continuum limit it
reproduces in $D<4$ the standard action of the scalar matrix field with quartic
self-interaction. I calculate explicitly the  correlator of two
gauge fields $U_\mu(x)$ and $U_\mu^\dagger(x)$ at the same link
$(x,\mu)$ which completely determines all other correlators of the model
and show that it might undergo phase transitions at the values of parameters
where the proper one-matrix model does.

\newsection{A review of the DMS approach}

Let us define for the Kazakov--Migdal model~\rf{spartition} the loop
average
\be
E_\l\equiv \LA\ntr{}\Big(\frac{1}{\l-\phi_x}\Big)\RA
\label{defE}
\ee
which describes all correlators of powers of $\phi_x$ at the same site
and the one-link correlator
\be
G_{\nu\lambda}\equiv\LA\ntr{}\Big(
\frac{1}{\nu-\phi_x}U_{\mu}(x)\frac{1}{\lambda-\phi_{x+\mu}}
U_\mu^\dagger(x) \Big) \RA \,.
\label{defG}
\ee
Both functions are analytic on the complex plane with some cut (or cuts)
at the real axis and obey the following asymptotic expansions
\be
E_\nu=\frac 1\nu + \sum_{k=1}^\infty \frac{{\cal E}_{k}}{\nu^{k+1}}~~,~~~~~~~
{\cal E}_k=\LA\ntr \phi^k_x \RA
\label{bc1}
\ee
and
\be
G_{\nu\l}=\frac{E_\l}{\nu} + \sum_{n=1}^\infty
\frac{G_n(\l)}{\nu^{n+1}} ~~,~~~~~~~
G_n(\l)=\LA\ntr{}\Big( \phi^n_x U_\mu(x)\frac{1}{\l-\phi_{x+\mu}}
U^\dagger_\mu(x) \Big)\RA \,.
\label{bcG}
\ee
Notice that
$G_{\nu\la}$ is symmetric in $\nu$ and $\la$
due to invariance of the Haar measure, $dU$, under
transformations $U\ra U^\dagger$.

The correlator $G_{\nu\l}$ obeys in the large-$N$
limit the following equation~\cite{DMS93}
\be
 \int_{C_1} \frac{d \om}{2\pi i}
\frac{\VVp(\om)}{\nu - \om}G_{\om \la}=
E_\nu G_{\nu \la} + \la G_{\nu \la} - E_\nu \, ,
\label{main}
\ee
where the contour $C_1$ encircles counterclockwise the cut (or cuts)
of the function $G_{\nu\la}$
(the same as the support of $\rho$)
and
\be
\VVp(\om)\equiv V^\prime(\om)-(2D-1) F(\om)\,.
\label{defL}
\ee

The function
\be
F(\om)=\sum_{n=0}^\infty F_n \om^n
\ee
 is  the one which appears in the expansion
of the pair correlator of the gauge fields
\be
\left\langle \frac {1}{N} \tr{} \Big(t^a U \chi
U^\dagger\Big)\right\rangle_{U} \equiv
\frac{\int dU\,\e^{N\tr{}\Big( \phi U
\chi U^\dagger\Big)} \frac {1}{N} \tr{} \Big(t^aU
\chi U^\dagger\Big)} {\int dU\,\e^{N
\tr{}\Big(\phi U \chi U^\dagger\Big)}}
\label{onelink}
\ee
where the averaging is only w.r.t.\ $U$ while $\phi$ and $\chi$ play
the role of external fields.  As was proposed in
Refs.~\cite{Mig92a,Mig92d}, the following formula holds at $N=\infty$:
\be
\left\langle \ntr{}\Big( t^a U \chi
U^\dagger
\Big)\right\rangle_{U} =\sum_{n=1}^\infty  F_{n}
\ntr{}\left(t^a\phi^{n}\right) \,,
\label{Lambda}
\ee
where $t^a$ ($a=1,\ldots,N^2-1$) stand for the generators of the $SU(N)$
which are normalized by
\be
\sum_a (t^a)^{ij} (t^a)^{kl} = N\delta^{il}\delta^{kj} -
\delta^{ij}\delta^{kl} \,.
\label{completeness}
\ee

The structure of the r.h.s.\ of \eq{Lambda} can be understood analyzing
the power series expansion of the Itzykson--Zuber correlator in
$\phi$ and $\chi$:
\be
\left\langle \ntr\Big({t^a U t^b U^\dagger}\Big) \right\rangle_U =
\sum_{n,m} {C}_{nm}
\ntr{}\Big(t^a \phi^n \Big)\ntr{}\Big(t^b \chi^m \Big)\,,
\label{Cnm}
\ee
where the coefficients $C_{nm}$ depend only on the distribution of
eigenvalues of $\phi$ and $\chi$. This formula is due to the invariance
of the Haar measure $dU$, which forbids terms like $\tr{\phi\chi}$,
as well as to the fact that only the
traceless parts of $\phi$ and $\chi$ contribute to the integral in
\eq{Cnm}.
Multiplying \eq{Cnm} by $\tr{t^b\chi}$ and using the completeness
condition~\rf{completeness}, one identifies $F_n$ with
\be
F_n=\sum_m {C}_{n,m} \Big( \ntr{\chi^{m+1}}-
\ntr{\chi}\ntr{\chi^m} \Big)~,~~~~~~~~\hbox{for \ \ } n\geq 1~.
\ee

Eq.~\rf{defL} arise after the multiplication of~\eq{Lambda} by
$\tr{}(t^a/(\nu-\phi))$ which gives
\be
\LA \ntr{} \Big(\frac{1}{\nu-\phi}U \chi U^\dagger \Big)\RA_U
=  \ntr{} \Big( \frac{F(\phi)}{\nu-\phi} \Big)
+\ntr \Big( \frac{1}{\nu-\phi} \Big) \ntr
\Big( \phi-F(\phi) \Big)\,.
\label{theansatz}
\ee
Since $F_0$ does not enter this formula, one can always choose
\be
F_0=\ntr{} \Big( \phi - \sum_{n=1}^\infty F_n \phi^n \Big)
\label{defF_0}
\ee
 to provide
\be
 \ntr \Big( \phi-F(\phi) \Big) = 0 \,.
\label{F_0}
\ee
 For the even  potential $V(\phi)=V(-\phi)$
\eq{F_0} holds automatically
since the distribution of eigenvalues is symmetric,
$\rho(\l)=\rho(-\l)$, and $F(\l)=-F(-\l)$.

With $F_0$ given by~\rf{defF_0}, \eq{theansatz} can be rewritten as
\be
G_1(\nu) = \int_{C_1} \frac{d \om}{2\pi i}
\frac{F(\om)}{\nu - \om} E_\om
\label{1}
\ee
where $G_1(\nu)$ is defined by the expansion~\rf{bcG}.
The same quantity appears in the simplest loop equation
(see~\cite{Mak92b} for a review)
\be
 \int_{C_1} \frac{d \om}{2\pi i}
\frac{{V}^\prime(\om)}{\nu - \om} E_\om -2D G_1(\nu) = E^2_\nu
\label{loopE}
\ee
which is just a result of the infinitesimal shift
\be
 \phi_x \ra \phi_x + \frac{\epsilon}{\nu-\phi_x}
\ee
of the measure in \rf{spartition} and can be alternatively obtained
taking the $1/\l$
term of the expansion of \eq{main} in $\l$.
Inserting~\rf{1} into \eq{loopE} and using~\rf{defL}, one arrives at the
following equation for $E_\nu$
\be
 \int_{C_1} \frac{d \om}{2\pi i}
\frac{\tilde{V}^\prime(\om)}{\nu - \om} E_\om  = E^2_\nu
\label{EE}
\ee
where
\be
\tilde{V}^\p(\l) = \VVp(\l) - F(\l) \,.
\label{tildeVp}
\ee

One can see now that \eq{F_0} can be rewritten using Eqs.~\rf{EE} and
\rf{tildeVp} as
\be
\ci \VVp(\om) E_\om = {\cal E}_1
\ee
which coincides with the $1/(\nu\l)$ term of the double expansion of \eq{main}
in $1/\nu$ and $1/\l$. Therefore, the choice~\rf{defF_0} is always
compatible with \eq{main}.

While \eq{EE} coincides with the large-$N$ loop equation for
the Hermitean one-matrix model, the potential $\tilde{V}(\om)$
is, generally speaking, non-polynomial and has singularities on the
complex plane outside of the cut (or cuts) of $E_\om$.
Another interesting property of the above formulas is that at $D=1/2$,
which is associated with the Hermitean two-matrix model, the last two
terms on the r.h.s.\  of \eq{defL} disappear and
one gets just ${\cal V}(\om)=V(\om)$. For this reason the
Schwinger--Dyson equations for the Hermitean two-matrix model~\cite{S,AK}
are equivalent to \eq{main}.

\newsection{Exact solution for arbitrary potential}

Let us consider the two-matrix model with an arbitrary potential
\be
{\cal  V}(\omega)=\sum_{m=1}^\infty \frac{g_{m}}{m} \omega^{m} \,.
\label{def L}
\ee
The solution of loop equations in this case was obtained in
Ref.~\cite{S} and in a more explicit form in Ref.~\cite{DMS93}.
The equation~\rf{main} for the potential~\rf{def L} reads explicitly
\be
\left(\lambda + E_\nu - \VVp(\nu) \right)G_{\nu\lambda}~=~E_\nu -R_\l(\nu)
\label{gen eq for G}
\ee
where $R_\l(\nu)$ is given by
\be
R_\l(\nu) \equiv -\cii \frac{\VVp(\om)}{\nu-\om} G_{\om\l} =
\sum_{m=2}^\infty g_{m}\sum_{n=0}^{m-2} G_n(\lambda)\nu^{m-n-2}
\label{defR}
\ee
and the contour $C_2$ encircles both singularities of $G_{\om\l}$ and
the pole at $\om=\nu$.
The  terms on the r.h.s.\ result from taking the
residue at infinity in the contour integral
and the functions $G_n(\l)$ are defined by \eq{bcG}.

The formal solution to \eq{gen eq for G} is
\be
G_{\nu\l} = \frac{E_\nu-R_\l(\nu)}{\l+E_\nu-\VVp(\nu)}
\label{genG}
\ee
with $R_\l(\nu)$ given by~\rf{defR}.
The functions $G_n(\l)$ can be expressed via $E_\l$ using the recurrence
relation
\be
G_{n+1}(\l)=\ci \frac{\VVp(\om)}{\l-\om} G_n(\om) - E_\l G_n(\l)~,
{}~~~~~~G_0(\l)=E_\l
\label{recurrent}
\ee
which can  be obtained expanding \eq{main} in $1/\l$.
For $n=0$ this equation recovers \eq{1}.

To determine $E_\l$, I use the equation~\cite{DMS93}
\be
\ci \,\VVp(\om) G_{\om\l} =  \l E_\l -1
\label{smart integral}
\ee
which is nothing but the $1/\nu$ term of the expansion of \eq{main}
in $1/\nu$. Using the expansion~\rf{bcG}, one can rewrite
\eq{smart integral}
as
\be
\sum_{m\geq1} g_{m}G_{m-1}(\l) = \l E_\l -1~.
\label{smart}
\ee
 For ${\cal V}(\l)$ being a polynomial of the highest power $J$,
\eq{smart} contains $E_\l$ up to the power $J$ and the solution is
therefore {\it algebraic}. This fact was first noted by
Staudacher~\cite{S} for the two-matrix model.
An explicit example of a non-polynomial potential which results in a quadratic
equation for $E_\l$ is  given below.

As is proven in Ref.~\cite{DMS93}:
\begin{itemize}
\item \vspace{-7pt}
 Equations which appear from
the next terms of the $1/\nu$-expansion of
\eq{main} are automatically
satisfied as a consequence of Eqs.~\rf{recurrent} and \rf{smart}.
\item \vspace{-6pt}
$G_{\nu\l}$ given by \rf{genG} is symmetric in $\nu$ and $\l$
for any solution of \eq{smart}.
The symmetry requirement can be used directly to determine $E_\l$
alternatively to \eq{smart}.
\end{itemize}
\vspace{-6pt}

It is worth mentioning the relation to the approach by Migdal~\cite{Mig92a}
 which is based on the Riemann--Hilbert method. As
is noticed in Ref.~\cite{DMS93}, \eq{main} is equivalent to the equation
of Ref.~\cite{Mig92a}. To show this let us define the continuous and
discontinuous in $\nu$ parts of $G_{\nu\l}$ across the cut (cuts) by
\bea
& &\Di G_{\nu\l} \equiv \frac{G_{(\nu+i0)\l}-G_{(\nu-i0)\l}}{2}~, \non
& &\Co G_{\nu\l} \equiv \frac{G_{(\nu+i0)\l}+G_{(\nu-i0)\l}}{2}
\eea
so that for a real $\l$ outside of the cut (cuts)
$\Di G_{\nu\l}$ coincides with
the imaginary part and $\Co G_{\nu\l}$ coincides with the real part.
In particular, $\Di E_\nu = -\pi \rho(\nu)$. The discontinuous part of
\eq{main} then reads
\be
\Co G_{\nu\l} = 1 + \frac{1}{\pi \rho(\nu)}
\Big( \l +\frac 12 \tilde{V}^\p(\nu) -\VVp(\nu) \Big)\Di G_{\nu\l}~,~~~~~~~
\hbox{for \ } \nu\in \hbox{ cut}
\label{ME}
\ee
which coincides with the equation of Ref.~\cite{Mig92a}.

To obtain the solution to \eq{ME} for $ G_{\nu\l}$ versus $E_\nu$,
one notices that for any real $\nu$
\be
\frac{1-G_{(\nu+i0)\l}}{1-G_{(\nu-i0)\l}}
=\frac{1-\Co G_{\nu\l}- i\Di G_{\nu\l} }
{1-\Co G_{\nu\l}+i\Di G_{\nu\l}}
= \frac{\l+ \real E_\nu - i \im E_\nu - \VVp(\nu)}
{\l+ \real E_\nu + i \im E_\nu - \VVp(\nu)}
\label{RGproblem}
\ee
since $\Di G_{\nu\l}$ cancels at the cut (cuts) due to \eq{ME}.
The analytic function which solves this problem
 is given by~\cite{Mig92a,Gro92,Bou93}
\be
G_{\nu\l} = 1- \exp{\left\lbrace
-\ci \frac{1}{\nu-\om}\log{(\l+E_\om-\VVp(\om)})\right\rbrace}~.
\label{RGsolution}
\ee
Taking the residue at $\om=\nu$, one can rewrite it in the
form~\rf{genG} with
\be
R_\l(\nu) =
\exp{\left\lbrace
-\cii \frac{1}{\nu-\om}\log{\left(1+\frac{E_\om}{\l-\VVp(\om)}
\right)}\right\rbrace}~.
\ee
The l.h.s.\ is obviously analytic in $\l$, expandable in $1/\l$,
has no discontinuity in $\nu$ and is expandable in powers of $\nu$ ---
\ie has all the properties required for $R_\l(\nu)$.

The formula~\rf{RGsolution} should solve the recurrence
relation~\rf{recurrent}.
I have checked this expanding in $1/\l$ to few lower orders.
Analogously, the master field equation
\be
E_\l = \ci \log{(\l+E_\om-\VVp(\om)})\,,
\label{MFE}
\ee
which is obtained as the $1/\nu$ term of \eq{RGsolution},
should be equivalent to \eq{smart}.
It is a matter of practical
convenience which form of equations to solve. As was shown in
Ref.~\cite{DMS93}, Eqs.~\rf{main} and \rf{smart} can  easily be
solved for the quadratic  potential ${\cal V(\phi)}$.
A simple solution of Eqs.~\rf{main}, \rf{smart} for the Penner
potential is obtained in the next section.

\newsection{An explicit solution for Penner potential}

Let us choose the following potential of the two matrix model
\be
{\cal V}(\phi) =  -(ab+c) \log{(b-\phi)}  - a \phi =
\sum_{m=2}^\infty \frac{ab+c}{m\,b^{m}} \phi^m + \frac cb \phi
\label{ttV}
\ee
where $a$ and $b$ are real. This potential recovers the quadratic one
in the limit
\be
a,b\ra\infty~,~~~~~~\frac ab\equiv \frac 1\L  \sim1~,~~~~~~c\sim1
\hbox{ \ \ \ \ (quadratic potential)}~.
\label{limit}
\ee

The loop equation~\rf{main} involves
\be
 \VVp(\om) =
\frac{a\om+c}{b-\om}\,.
\label{penL}
\ee
The function $R_\l(\nu)$ which enters \rf{genG} can then be calculated
to give
\be
R_\l(\nu) \equiv -\cii \frac{\VVp(\om)}{\nu-\om} G_{\om\l}
= \frac{ab+c}{b-\nu} G_{b\l}\,.
\label{penR}
\ee
This formula is obtained doing the integration over
 the contour $C_2$ which encircles both singularities of $G_{\om\l}$ and
the pole at $\om=\nu$. For $\VVp(\om)$ given by~\rf{penL} the residue
at infinity vanishes since $G_{\om\l}$ decreases as $1/\om$ while the
residue at $\om=b$ results in the expression on the r.h.s.. One can
alternatively derive this formula expanding $\VVp(\om)$ in $\om$ and
taking the residue at infinity.

The function $G_{b\l}$ on the r.h.s.\ of \eq{penR} can be determined
from \eq{smart integral} to be
\be
G_{b\l}=\frac{(a+\l) E_\l -1}{ab+c}
\label{Gbl}
\ee
so that the solution to \eq{main} for $G_{\nu\l}$ versus $E_\l$ is
\be
G_{\nu\l} = \frac{E_\nu-
\frac{(a+\l) E_\l -1}{b-\nu}}{\l+E_\nu-\frac{a\nu+c}{b-\nu}} \,.
\label{penG}
\ee

To determine $E_\l$ let us utilize the symmetry of $G_{\nu\l}$ in $\nu$
and $\l$. From \eq{penG} one gets
\be
G_{\nu b}=\frac{(b-\nu) E_\nu - (a+b)E_b +1}{(b-\nu)(b+E_\nu)-a\nu-c }
\begin{array}[b]{c}\hbox{\footnotesize (\ref{Gbl})} \\ = \end{array}
\frac{(a+\nu) E_\nu -1}{ab+c}
\ee
which yields the {\it quadratic}\/ equation for $E_\nu$
\bea
(a+\nu)(b-\nu)E_\nu^2 - \Big((a\nu+c)(a+\nu)+(1+c-b\nu)(b-\nu) \Big)
E_\nu \non
=-\nu(a+b)+b^2-c - (ab+c)\Big((a+b) E_b-1\Big)  \,.
\label{quaE}
\eea
I have verified that~\rf{penG} is indeed symmetric in
$\nu$ and $\l$ providing \eq{quaE} is satisfied. The formula for
$C(\nu,\l)$, the double discontinuity of $G_{\nu\l}$ both in $\nu$ and
$\l$ across the cut (cuts), which is manifestly symmetric in $\nu$ and $\l$ is
presented in Sect.~\ref{phase}.

\eq{quaE} exactly coincides with the large-$N$ loop equation for the
Hermitean one-matrix model with the logarithmic potential
\be
\tilde{V}(\phi)= -(ab+c)\log{(b-\phi)}+ (ab+c+1) \log{(a+\phi)} -(a+b)\phi
\label{tV}
\ee
for which
\be
\tilde{V}^\p(\l) = \frac{a\l+c}{b-\l} +\frac{1+c-b\l}{a+\l} \,.
\label{tVp}
\ee
The constant $E_b$ is a free parameter which should be fixed requiring
the analytic structure of $E_\l$ on the complex $\l$-plane.

It it easy to demonstrate that the solution~\rf{penG} satisfies \eq{ME}.
To show this let us use the symmetry of~\rf{penG} in $\nu$ and $\l$ and
calculate $\Di G_{\l\nu}$ and $\Co G_{\l\nu}$ which read
\be
\frac{1}{\pi\rho(\nu)}\Di G_{\l\nu} = \frac{\frac{a+\nu}{b-\l}}
{\nu+E_\l-\frac{a\l+c}{b-\l}}~,~~~~~~
\Co G_{\l\nu} = \frac{E_\l-\frac{\half (a+\nu)\tilde{V}^\p(\nu)-1}{b-\l}}
{\nu+E_\l-\frac{a\l+c}{b-\l}} \,.
\ee
These expressions identically satisfy \eq{ME} and, therefore,
\eq{RGproblem}. I conclude that \rf{penG}
solves the Riemenn-Hilbert problem and should coincide
with~\rf{RGsolution} for the potential $\VVp$ given by \eq{penL}.
Analogously, \eq{MFE} should reduce to \eq{quaE}.

It is instructive to discuss the one-cut solution to \eq{quaE} which
reads~\cite{Mig83}
\be
E_\l = \int_{C_1} \frac{d\om}{4\pi i} \frac{\tilde{V}^\p(\om)}{\l-\om}
\frac{\sqrt{(\l-x)(\l-y)}}{\sqrt{(\om-x)(\om-y)}}
\label{EvstV}
\ee
where the ends of the cut, $x$ and $y$, are determined by the asymptotic
conditions
\be
\ci \frac{\tilde{V}^\p(\om)}{\sqrt{(\om-x)(\om-y)}} =0\,~~~~
\ci \frac{\om \tilde{V}^\p(\om)}{\sqrt{(\om-x)(\om-y)}} =2 \,.
\label{xandy}
\ee
For $\tilde{V}^\p$ given by~\rf{tVp} the contour integral can easily
be calculated taking the residues at $\om=\l,b$ and $-a$ while the residue at
infinity vanishes since $E_\om$ falls down as $1/\om$.  One gets
\bea
E_\l = \frac 12 \left( \frac{a\l+c}{b-\l} + \frac{1+c-b\l}{a+\l}\right) -
  \frac 12 \left( \frac{ab+c}{b-\l} \frac{1}{\sqrt{(b-x)(b-y)}}\right. \non
+ \left. \frac{ab+c+1}{a+\l}\frac{1}{\sqrt{(-a-x)(-a-y)}}\right)
\sqrt{(\l-x)(\l-y)}  \,.
\label{one-cut}
\eea

It is worth noting that all the formulas recover the ones for the
Gaussian potential in the limit~\rf{limit}. Since in the Gaussian case
\eq{xandy} has the solution
\be
y=-x,~~~~~~~ x^2 =\frac ab  - \frac ba~,
{}~~~~~~~\hbox{ (quadratic potential)}~,
\ee
the one-cut
solution~\rf{one-cut} is always realized for $a$ and $b$ which are big
enough for the points $b$ and $-a$ to lie outside of the cut $[x,y]$.

The situation when ether $b$ or $-a$ touch the end of the cut is associated
with a phase transition to a more-than-one-cut solution.
The one-matrix model with the potential~\rf{tV} has at least as rich
structure as that with the cubic potential since the latter can be
obtained from~\rf{tV} taking the limit
\be
a,b\ra \infty,~~~~(a-b)\sim b^{\frac 13}~~~~~~~~~\hbox{(cubic potential)}
\label{cubic potential}
\ee
and rescaling the field $\phi\ra\phi b^{\frac 13}$ which can always be
done for the one-matrix model. The linear in $\phi$  term is finite
providing $c\laa b^{\frac 23}$. It can be explicitly seen that in the
limit~\rf{cubic potential} the one-cut~\rf{one-cut}
solution recovers the one for the cubic potential.

Since $\tilde{V}'(\l)$ is known, the function $F(\l)$ can be determined
from \eq{tildeVp} to be
\be
F(\l) = \frac{b\l-c-1}{a+\l} \,.
\label{penLambda}
\ee
Now \eq{defL} determines
\be
V^\p(\l) = \frac{a\l+c}{b-\l} + (2D-1) \frac{b\l-c-1}{a+\l}
\ee
which corresponds to the potential
\be
V(\phi)= -(ab+c) \log{(b-\phi)}- (2D-1) (ab+c+1) \log{(a+\phi)}
+[(2D-1)b-a] \phi \,.
\label{V}
\ee
This formula recovers at $D=1/2$ the potential~\rf{ttV} of the two-matrix
model and at $D=0$ the potential~\rf{tV} of the associated one-matrix
model.
In the Gaussian limit~\rf{limit} the solution found by Gross~\cite{Gro92}
is reproduced.
Analogously, I verified that the next term of the $1/b$ expansion
recovers a perturbation of the model with the quadratic
potential by a cubic term with the
coupling $g_3=a/b^2$.

\newsection{The ``naive'' continuum limit}

The potential~\rf{V} of the $D$-dimensional Kazakov--Migdal model admits
the following ``naive'' continuum limit. Let us expand $V$ in $\phi$
which yields
\bea
V(\phi)=\left( \frac{c}{b} -(2D-1)\frac{c+1}{a} \right) \phi
+ \frac 12 \left( \frac{ab+c}{b^2}+(2D-1)\frac{ ab+c+1}{a^2} \right) \phi^2
\non
+ \frac 13 \left(\frac{ab+c}{b^3}-(2D-1)\frac{ ab+c+1}{a^3} \right)
\phi^3  \non
+ \frac 14 \left( \frac{ab+c}{b^4}+(2D-1)\frac{ ab+c+1}{a^4} \right)
\phi^4+ {\cal O} (\phi^5)\,.
\label{expansion}
\eea
Now the idea is to choose the couplings $a,b$ and $c$ to provide the
canonical scaling
\be
\phi=\eps^{\frac D2 -1}\Phi~,~~~~~~~~V(\phi)=\eps^{D}W(\Phi)
\label{canonical}
\ee
which is prescribed, as usual, by the kinetic term.
The expansion will be justified choosing $a\sim b\ra\infty$
so that $\phi/b\ra0$.
The linear in $\phi$ term has the continuum limit providing
\be
\left(\frac ab +1-2D\right)c-(2D-1)\laa \eps^{\frac D2+1} a
\label{eq for c}
\ee
which can always be satisfied by a proper choice of $c$.

The bare mass
\be
m_0^2 = \frac ab + (2D-1)\frac ba  \geq 2D
\ee
for any $D$ and the equality sign holds only for $D=1$ at $a/b=1$.
Since the equality is needed to cancel the term $2D\phi^2$ coming from the
kinetic energy, one concludes that the ``naive'' continuum limit is
possible only for $D=1$.

At $D=1$ one has
\be
m_0^2=2+\frac{(a-b)^2}{ab} = 2+\eps^2 M^2
\ee
where
\be
\frac{a-b}{\eps b} = M \sim1
\ee
is the continuum mass.
Now the cubic in $\phi$ term has the continuum
limit providing
\be
a=b=\sqrt{\frac {2}{g}}\eps^{\frac D2 - 2}
\label{bcubic}
\ee
which justifies the expansion in
\be
\frac{\phi}{b} \sim \eps \,.
\ee
Notice that the quartic in $\phi$ term is also finite for the
$\eps$-dependence of $b$ given by \eq{bcubic} as well as \eq{eq for c}
is satisfied for any $c\sim1$.
We arrive, therefore, at
the  one-dimensional continuum quartic action
\be
S[\Phi] = \int dt \Big( \hbox{const.} \Phi(t) + \frac 12 \dot{\Phi}^2(t)
+ \frac{M^2}{2} \Phi^2(t) +
\sqrt{\frac g2} M \Phi^3(t) + \frac {g}{4} \Phi^4(t)\Big)
\label{D=1}
\ee
where $\dot{\Phi}\equiv \partial_t \Phi$.
The coefficient in front of the linear in $\phi$ term can be made
arbitrary (or vanishing) by choosing $c$.

The above procedure of taking the ``naive'' continuum limit
works for $D<4$ where $b$ given by \eq{bcubic} is divergent. The only
difference is that for $D\neq1$ the bare mass term remains infinite.
One can cancel it, however, for $2\leq D<4$ by the standard
{\it renormalization} procedure. The restriction $D<4$ is precisely the one
where the scalar theory with the quartic interaction is renormalizable.

As was shown by Gross~\cite{Gro92}, any solution of the master field
equation~\rf{MFE} for an arbitrary potential
reproduces in $D=1$ the spectral density given by the fermionic solution of
Ref.~\cite{BIPZ78}. For this reason there is no doubt that the solution
of the previous section does. However, an interesting question is   how
the one-cut solution~\rf{one-cut} can reproduce this $D=1$ spectral
density? The answer is that the potential~\rf{D=1} is simple enough to
provide at $D=1$ the factorized structure
\be
\pi \rho(\l)=\sqrt{2E_F -2 W(\l)}= \sqrt{\frac g2}(\l+ \hbox{const.})
\sqrt{(y-\l)(\l-x)}
\ee
by turning the coupling in front of the linear in $\phi$ term.

\newsection{The Itzykson--Zuber correlator \label{phase}}

To calculate the pair correlator of $U$ and $U^\dagger$, one should
take the discontinuity of~\rf{penG} both in $\nu$ and $\l$
across the cut (cuts)~\cite{DMS93}. For the solution~\rf{penG} one gets
\be
C(\nu,\l) \equiv \frac{1}{\pi^2 \rho(\nu) \rho(\l)}\,\hbox{Disc}_\l\,
\Di G_{\nu\l}=
\frac{(a+\nu)(a+\l)}{{\cal D}(\nu,\l)}~,~~~~~~~~~\nu,\l\in \hbox{cut}
\label{CC}
\ee
 with
\bea
{\cal D}(\nu,\l)=-\l^2\nu^2+(b-a)\l\nu(\l+\nu)
+ab(\l^2+\nu^2)-
\l\nu(a^2+b^2+2c+1) \non +(\l+\nu)(b-ac+bc)-c^2-b^2
+(ab+c)[(a+b)E_b-1]\,,
\label{CCC}
\eea
where $E_b$ depends on the type of the solution of \eq{quaE}
(one-cut or more-than-one-cut solutions). In the Gaussian
limit~\rf{limit} when $E_b\ra 1/b$, one recovers the result of
Ref.~\cite{DMS93}.

While the expression~\rf{CC} looks complicated,
the denominator seems to be positive.
To see this let us rewrite ${\cal D}(\nu,\l)$ as
\be
{\cal D}(\nu,\l) = (b-\nu)(a+\nu) (\l-r_+(\nu))(\l-r_-(\nu))
\ee
and
\be
r_\pm(\nu)=\half (F(\nu)+\VVp(\nu)) \pm i \pi \rho(\nu) \,.
\label{roots}
\ee
Since $\rho(\nu)$ is real for $\nu \in$ cut, the roots are complex
except for the values of parameters when $\rho(\nu)$ vanishes for some
 $\nu\in$ cut.
This might happen for
 the cases of  phase transitions in the one-matrix model with the
potential~\rf{tV}. The simplest one is from the one-cut
solution to a two-cut solution. It is an algebraic problem to study
whether the roots~\rf{roots} always lie outside of the cut (cuts).

These phase transitions does not change, however, the functional
structure of $C(\nu,\l)$ given by Eqs.~\rf{CC}, \rf{CCC} --- only the
constant $E_b$ changes.
For this reason the Kazakov--Migdal model with the logarithmic
potential~\rf{V} seems to be too simple to exhibit a non-trivial critical
behavior at $D>1$.

\subsection*{Acknowledgements}
I am grateful to J.~Ambj{\o}rn,
D.~Boulatov, M.~Dobroliubov, C.~Kristjansen  and G.~Semenoff for
useful discussions. I thank the
NBI high energy theory group for the hospitality at Copenhagen.


\end{document}